# Permanent magnet with MgB$_2$ bulk superconductor

Akiyasu Yamamoto*[1, 3], Atsushi Ishihara[2], Masaru Tomita[2] & Kohji Kishio[1]

[1]*The University of Tokyo, 7-3-1 Hongo, Bunkyo, Tokyo 113-8656, Japan*

[2]*Railway Technical Research Institute, 2-8-38 Hikari, Kokubunji, Tokyo 185-8540, Japan*

[3] *JST-PRESTO, 4-1-8 Honcho, Kawaguchi, Saitama 332-0012, Japan*

**Abstract:**

Superconductors with persistent zero-resistance currents serve as permanent magnets for high-field applications requiring a strong and stable magnetic field, such as magnetic resonance imaging (MRI). The recent global helium shortage has quickened research into high-temperature superconductors (HTSs)—materials that can be used without conventional liquid-helium cooling to 4.2 K. Herein, we demonstrate that 40-K-class metallic HTS magnesium diboride (MgB$_2$) makes an excellent permanent bulk magnet, maintaining 3 T at 20 K for 1 week with an extremely high stability (<0.1 ppm/h). The magnetic field trapped in this magnet is uniformly distributed, as for single-crystalline neodymium-iron-boron. Magnetic hysteresis loop of the MgB$_2$ permanent bulk magnet was detrmined. Because MgB$_2$ is a simple-binary-line compound that does not contain rare-earth metals, polycrystalline bulk material can be industrially fabricated at low cost and with high yield to serve as strong magnets that are compatible with conventional compact cryocoolers, making MgB$_2$ bulks promising for the next generation of Tesla-class permanent-magnet applications.

Superconducting bulk magnets are a direct manifestation of quantum phenomena on the macroscopic scale. By magnetizing at below its transition temperature $T_c$, the entire bulk acts as a compact Tesla-class magnet because of the presence of induced macroscopic circulating supercurrents [1-8]. If the magnetized bulk is kept cold, it will act as a permanent magnet. Superconducting bulk magnets are suitable for new compact-magnet applications such as desktop NMR, MRI, motors, and particle accelerators [9-11], which require very strong fields that cannot be obtained by conventional permanent magnets and flexible magnet-shape design. State-of-the-art texturing growth techniques overcome the electromagnetic weak-link problem in quasi-single-crystal forms of HTS rare-earth barium copper oxide (HTS-REBCO), and a record high field of 17 T was trapped in a 1-inch compact bulk [2,8]. Although various potential applications of superconducting bulk magnets have emerged [12,13], insufficient field uniformity and reproducibility due to anisotropic and unpredictable crystal growth hinder the industrial use of HTS-REBCO bulks. Efforts to improve their properties, especially field uniformity [14-16] and productivity [17,18], are being performed.

$MgB_2$ is a metallic HTS discovered in 2001 [19]. Because of its high $T_c$ (40 K) and recent progress in high-performance cryocoolers, it is a strong contender for future cryocooled, liquid-helium-free applications at 5–30 K. In practice, the distinct advantage of $MgB_2$ is that its metallic superconductivity (i.e., it's simple and symmetric pairing mechanism, high carrier density, and long coherence length) drives a weak-link-free supercurrent flow across grain boundaries, and this occurs even in untextured polycrystalline bulks [20]. Moreover, the nanoscale grain boundaries strongly pin the flux quantum [21-24], and in superconducting bulk magnets, this effect is expected to yield strong, uniform, and stable magnetic fields. We recently found that a $MgB_2$ bulk with a large trapped magnetic field can be synthesized by simple sintering; a technique that can lead to the low-cost, high-yield, and scalable production of superpowerful liquid-helium-free magnets.

Bulk polycrystalline $MgB_2$ was synthesized by a conventional sintering process with a mixture of magnesium and boron powders and by using a modified powder-in-closed-tube technique [25]. Magnesium (99.8% purity) and boron (99.9% purity) powders were weighed and then mixed at a molar ratio of 1:2 in an agate mortar. Subsequently, the resulting powder was uniaxially pressed at 100 MPa to form a disk-shaped pellet (30 mm in diameter and 10 mm thick). To prevent the magnesium from oxidizing, the pellet was sealed in a stainless-steel tube and heat treated in a tube furnace at 850 °C for 3 h under an argon atmosphere.

The inset of Fig. 2 shows the photograph of a disk-shaped bulk $MgB_2$. The surface structure is homogeneous with no cracks or crystal domains to disrupt the uniform supercurrent flow. Analysis by powder X-ray diffraction confirms that the bulk is nearly-single-phase $MgB_2$ with small amount of MgO. Moreover, our bulk $MgB_2$ has a mass density of 1.3 g/cm$^3$, which is ~5.2 times lighter than REBCO (6.7 g/cm$^3$) and ~5.7 times lighter than neodymium–iron–boron (NIB: 7.4 g/cm$^3$) bulks.

To measure the magnetic flux density, two disk-shaped bulk $MgB_2$ (30 mm diameter and 10 mm thick) were stacked vertically across a spacer containing a transversal cryogenic Hall sensors (HGCT-3020, Lake Shore). The disk pair was cooled by a Gifford–McMahon (GM) cryocooler (CRT-HE05-CSFM, Iwatani Gas).

For practical applications of superconducting bulk materials as permanent magnets, one must determine the magnetic response and stability over time, including the change in the applied external field during the initial magnetization process and the residual magnetization as well as its corresponding creep as a function of time after field removal. To this end, we measured the cyclic magnetic hysteresis of the pair of $MgB_2$ bulk disks. After cooling the sample to various temperatures in the absence of a magnetic field, an external field was cyclically applied at a constant rate of ±1.8 T/h; the magnetization was recorded as the local magnetic flux density at the center of the sample. Hysteresis loops were obtained at 10, 20, and 30 K. In Fig. 1, $H_p$, $B_r$, and $-H_c$ represent the most important parameters describing the loops. The *penetration field* $H_p$ is the external field when the penetrating magnetic field front reaches the center of the sample disk. The quantity $B_r$ corresponds to the remnant local magnetization at the zero applied field of the descending step and is commonly referred to as the trapped *remnant field*. Similarly, $-H_c$ is the *coercive field* where the field in the magnet reaches zero. These practically important parameters in the macroscopic magnetization curve for $MgB_2$ can be defined in a similar manner to those for a conventional, electron spin-based permanent ferromagnets, even though the microscopic origin of the magnetism is entirely different for the bulk $MgB_2$ superconductor.

Figure 2 shows the temperature dependence of the magnetic field trapped in the pair of bulk $MgB_2$ disks. The sample was cooled to ~10 K under an external field of 4.5 T, following which the external field was removed. After field-cooling magnetization, the magnetic flux density trapped in the disk pair was measured at two positions, including at the center of the spacer (bulk center) and at the center of the bulk surface (bulk surface) as a function of temperature at a sweep rate of 0.1 K/min. A maximum trapped field of 4.02 T at 11 K was recorded, which is three times larger than the remnant magnetization $B_r$ of conventional NIB materials. With increasing temperature, the field decreased continuously due to a decreasing critical current density in the bulk, until it vanished at 38 K ($T_c$ for $MgB_2$). At 15, 20, 25, and 30 K the trapped field was 3.6, 2.9, 2.1, and 1.3 T, respectively. The difference in trapped fields at the surface and the center is due to a geometrical factor, and the ratio $B_{center}/B_{surface}$ ~ 1.35 is considered reasonable if we compare with Biot-Savart calculation (~1.6) which assumes that the current density is constant in bulk. These results suggest that $MgB_2$ bulks hold promise for applications requiring high-field magnets that function without liquid-helium cooling, such as 200 MHz NMR (4.7 T) at ~10 K and 1.5- to 3-T-class MRI at ~20 K.

Figure 3 shows the distribution of the local trapped field of a magnetized $MgB_2$ disk at 20 K. The data was obtained by scanning a Hall sensor 3 mm above the bulk surface. As shown in Fig. 3(a), the spatial distribution of the trapped field is conical, as expected for an ideal superconducting magnetized disk with a uniform supercurrent $J = 1/\mu_0 \times \partial B/\partial r$ [26]. The local field, as a function of position, was consistent between different radial directions (Fig. 3(b)), which suggests that the supercurrent is circumferentially perfectly homogeneous. As a result, contour lines for the equivalent local field (Fig. 3(c)) are circular (with a 99.4% degree of circularity), indicating that the trapped field is circumferentially constant, similar to that of a NIB disk. We attribute the excellent field uniformity to the metallic superconductivity of polycrystalline $MgB_2$. More precisely, the field uniformity is due to the weak-link-free uniform supercurrent flow and strong, uniform flux pinning, which in turn are caused by the naturally distributed nanoscale grain boundaries.

To assess the stability of bulk-$MgB_2$ magnets, the magnet creep was measured for 7 days. Figure 4(a) shows the time dependence of the trapped field in the 30-mm pair of bulk-$MgB_2$ disks. After magnetization at 20 K and holding at 20 K, the trapped field undergoes continuous decay. The trapped field after magnetization was 2.87 T and decreased to 2.82 T after 3 days, which corresponds to a loss of 1.7%. After 1 hour, the trapped field decreased linearly over time on a logarithmic scale and was well fit by $B/B(t = 0) = 1.018 − 2.84 \times 10^{-3} \ln(t)$ for $t > 10^4$, where $t$ is dimensionless time standardized by second. When the sample was magnetized and held at a higher temperature the decay rate of the trapped field increased and, at 30 K, was fit by $B/B(t = 0) = 1.027 − 4.06 \times 10^{-3} \ln(t)$. These results suggest that magnet creep is due to the motion of trapped flux quantum [27,28]. Flux quantum in pinning potential $U$ are depinned by the thermal activation process driven by the Lorentz force $F_L = \Phi_0 \times J$ at the rate $R = \omega_0 \exp(-U/k_BT)$, where $\Phi_0$ is the flux quantum, $\omega_0$ is the attempt frequency of the flux, and $k_BT$ is the thermal energy.

The thermal activation of flux can be suppressed by operating at a temperature lower than that used to establish the critical state [29,30]. To further improve the magnet stability, the bulk pair was maintained at 19 K, which is 1 K below the magnetizing temperature. By lowering the holding temperature, the stability of the trapped field improved significantly, showing a decay of less than $1 \times 10^{-4}$ T in 1 week (Fig. 4(b)). The decay rate was fitted to $B_T(t)/B_T(0) = 1.000 − 4.79 \times 10^{-6} \ln(t)$, confirming the extremely slow average-magnet-creep rate of <0.1 ppm/h. This creep rate is identical to that specified for MRI scanners. Thus, the $MgB_2$ bulk magnet is a quasi-permanent magnet that is extremely stable at operating temperatures. We attribute the improved magnetic stability at 19 K to the decreased motion of trapped flux quantum, which causes the dissipation of circulating supercurrent. The flux is shifted upon transitioning from the creep state to the frozen state due to a decrease in $J/J_c$ at the operating temperature that increases the activation energy $U$ and suppresses thermal activation.

The trapped field as a function of temperature, position, time, and external field clearly shows that bulk polycrystalline $MgB_2$ is a novel strong Tesla-class quasi-permanent magnet that can produce a magnetic field greater than 4 T with excellent field uniformity, very high magnet stability, and a response to cyclic fields similar to that of spin-based magnets. The mechanism by which the high, uniform, and stable field in bulk $MgB_2$ develops is unique among HTS materials and is attributed to grain boundaries that act as homogeneous flux pinning sites rather than as current-blocking defects. On the other hand, iron and/or rare-earth elements, which are considered essential for spin-based permanent magnets, are not critical for superconducting bulk magnets because their high magnetic fields originate from macroscopic electromagnetic induction due to a persistently circulating supercurrent. Therefore, bulk liquid-helium-free superconducting quasipermanent magnets using $MgB_2$ or even new HTS materials are very interesting prospects in the ongoing search for strong magnets that do not require rare-earth metals [31].

Thus, our present results demonstrate that polycrystalline bulk superconducting $MgB_2$ is the strongest reported rare-earth-metal–free permanent magnet capable of operating with existing cryocoolers without conventional liquid-helium cooling. Given that the irreversibility field is greater than 10 T at 20 K [32], we expect that it is possible to trap even higher field by increasing the diameter of bulk $MgB_2$ and the current density, which can be done by tuning the electronic- and micro-structure [33,34] or by hybridizing with other HTSs. Because $MgB_2$ is free of rare-earth metals and is a simple binary line-compound, magnets made of bulk polycrystalline $MgB_2$ for use with compact cryocoolers can be industrially fabricated at low cost and with high yield. These make bulk $MgB_2$ extremely promising for the next generation of Tesla-class permanent-magnet applications, liquid-helium-free 3-T-class MRI, desktop NMR, and particle accelerators.

We are very grateful to T. Akasaka of RTRI and K. Iwase, J. Shimoyama of the University of Tokyo for experimental assistance. We thank E. E. Hellstrom and D. C. Larbalestier of the Applied Superconductivity Center, National High Magnetic Field Laboratory for their valuable suggestions on the manuscript and encouragement. This work was partially supported by a Grant-in-Aid for Scientific Research from the Japan Society for the Promotion of Science under grant Nos. 23246110 and 22860019 and by the Japan Science and Technology Agency, PRESTO.

**References**

[1] G. Fuchs, G. Krabbes, P. Schätzle, S. Gruβ, P. Stoye, T. Staiger, K.-H. Müller, J. Fink, and L. Schultz, *Appl. Phys. Lett.* **76**, 2107 (2000).

[2] M. Tomita, and M. Murakami, *Nature* **421**, 517 (2003).

[3] N. H. Babu, Y. Shi, K. Iida, and D. A. Cardwell, *Nat. Mater.* **4,** 476 (2005).

[4] A. Yamamoto, H. Yumoto, J. Shimoyama, K. Kishio, A. Ishihara, and M. Tomita, 23th Intl. Symp. on Superconductivity, Tokyo (November, 2010).

[5] T. Naito, T. Sasaki, and H. Fujishiro, *Supercond. Sci. Technol.* **25**, 095012 (2012).

[6] J. H. Durrell, C. E. J. Dancer, A. Dennis, Y. Shi, Z. Xu, A. M. Campbell, N. Hari Babu, R. I. Todd, C. R. M. Grovenor, and D. A. Cardwell, *Supercond. Sci. Technol.* **25**, 112002 (2012).

[7] G. Fuchs, W. Haessler, K. Nenkov, J. Scheiter, O. Perner, A. Handstein, T. Kanai, L. Schultz, and B. Holzapfel, *Supercond. Sci. Technol.* **26**, 122002 (2013).

[8] J. Durrell, A. Dennis, J. Jaroszynski, M. Ainslie, K. Palmer, Y. Shi, A. Campbell, J. Hull, M. Strasik, E. Hellstrom, and D. Cardwell, *Supercond. Sci. Technol.* **27**, 082001 (2014).

[9] Y. Iwasa, S. Y. Hahn, M. Tomita, H. Lee, and J. Bascuñán, *IEEE Trans. Appl. Supercond.* **15**, 2352 (2005).

[10] T. Nakamura, Y. Itoh, M. Yoshikawa, T. Oka, and J. Uzawa, *Concept Magn. Reson.* **B 31**, 65 (2007).

[11] T. Kii, R. Kinjo, N. Kimura, M. Shibata, M. A. Bakr, Y. W. Choi, M. Omer, K. Yoshida, K. Ishida, T. Komai, K. Shimahashi, T. Sonobe, H. Zen, K. Masuda, and H. Ohgaki, *IEEE Trans. Appl. Supercond.* **22**, 4100904 (2012).

[12] D. A. Cardwell, and N. H. Babu, *Physica C* **445-448**, 1 (2006).

[13] T. Oka, *Physica C* **463-465**, 7 (2007).

[14] M. Sekino, H. Yasuda, A. Miyazoe, and H. Ohsaki, *IEEE Trans. Appl. Supercond.* **21**, 1588 (2011).

[15] S. Hahn, J. Voccio, D. K. Park, K.-M. Kim, M. Tomita, J. Bascuñán, and Y. Iwasa, *IEEE Trans. Appl. Supercond.* **22**, 4302204 (2012).

[16] S. B. Kim, T. Kimoto, M. Imai, Y. Yano, J. H. Joo, S. Hahn, Y. Iwasa, and M. Tomita, *Physica C* **471**, 1454 (2011).

[17] M. Muralidhar, K. Suzuki, A. Ishihara, M. Jirsa, Y. Fukumoto, and M. Tomita, *Supercond. Sci. Technol.* **23**, 124003 (2010).

[18] Y. Shi, N. Hari Babu, K. Iida, W. K. Yeoh, A. R. Dennis, S. K. Pathak, and D. A. Cardwell, *Physica C* **470**, 685 (2010).

[19] J. Nagamatsu, N. Nakagawa, T. Muranaka, Y. Zenitani, and J. Akimitsu, *Nature* **410**, 63 (2001).

[20] D. C. Larbalestier, L. D. Cooley, M. O. Rikel, A. A. Polyanskii, J. Jiang, S. Patnaik, X. Y. Cai, D. M. Feldmann, A. Gurevich, A. A. Squitieri, M. T. Naus, C. B. Eom, E. E. Hellstrom, R. J.

Cava, K. A. Regan, N. Rogado, M. A. Hayward, T. He, J. S. Slusky, P. Khalifah, K. Inumaru, and M. Haas, *Nature* **410**, 186 (2001).

[21] P. C. Canfield, D. K. Finnemore, S. L. Bud'ko, J. E. Ostenson, G. Lapertot, C. E. Cunningham, and C. Petrovic, *Phys. Rev. Lett.* **86**, 2423 (2001).

[22] G. Giunchi, G. Ripamonti, T. Cavallin, and E. Bassani, *Cryogenics* **46**, 237 (2006).

[23] P. Mikheenko, E. Martínez, A. Bevan, J. S. Abell, and J. L. MacManus-Driscoll, *Supercond. Sci. Technol.* **20**, S264 (2007).

[24] T. Matsushita, M. Kiuchi, A. Yamamoto, J. Shimoyama, and K. Kishio, *Supercond. Sci. Technol.* **21**, 015008 (2008).

[25] A. Yamamoto, J. Shimoyama, S. Ueda, Y. Katsura, S. Horii, and K. Kishio, *Supercond. Sci. Technol.* **20**, 658 (2007).

[26] C. P. Bean, *Rev. Mod. Phys.* **36**, 31 (1964).

[27] P. W. Anderson, and Y. B. Kim, *Rev. Mod. Phys.* **36**, 39 (1964).

[28] T. Matsushita, *Flux Pinning in Superconductors* (Springer, German, 2006).

[29] M. R. Beasley, R. Labusch, and W. W. Webb, *Phys. Rev.* **181**, 682 (1969).

[30] G. Krabbes, G. Fuchs, W. R. Canders, H. May, and R. Palka, *High Temperature Superconductor Bulk Materials*, p.88 (Willey-VCH, German, 2006).

[31] N. Jones, *Nature* **472**, 22 (2011).

[32] A. Gurevich, *Nat. Mater.* **10**, 255 (2011).

[33] A. Yamamoto, J. Shimoyama, S. Ueda, Y. Katsura, I. Iwayama, S. Horii, and K. Kishio, *Appl. Phys. Lett.* **86**, 212502 (2005).

[34] V. Braccini, A. Gurevich, J.E. Giencke, M.C. Jewell, C.B. Eom, D.C. Larbalestier, A. Pogrebnyakov, Y. Cui, B. T. Liu, Y. F. Hu, J. M. Redwing, Q. Li, X. X. Xi, R. K. Singh, R. Gandikota, J. Kim, B. Wilkens, N. Newman, J. Rowell, B. Moeckly, V. Ferrando, C. Tarantini, D. Marré, M. Putti, C. Ferdeghini, R. Vaglio, and E. Haanappel, *Phys. Rev. B.* **71**, 012504 (2005).

**Figure captions:**

**Fig. 1.** (Color online)  Magnetic hysteresis loops of bulk-MgB$_2$ magnet at 10, 20, and 30 K. The magnetic field was measured at the center of the bulk-MgB$_2$ disk pair. Arrows indicate the counterclockwise direction of the loop. $H_p$, $B_r$, and $-H_c$ represent the penetration field, the trapped remnant field, and the coercive field, respectively.

**Fig. 2.** (Color online)  Trapped magnetic field measured as a function of continuously increasing temperature at center and surface of magnetized bulk-MgB$_2$ disk pair. The upper right inset shows schematic cross-sectional configuration of the bulk pair: two MgB$_2$ bulk disks and two Hall sensors. The lower-left inset shows the photograph of a bulk-MgB$_2$ disk.

**Fig. 3.** (Color online)  Distribution of trapped magnetic field in MgB$_2$ bulk disk magnet at 20 K measured 3 mm above bulk surface. (a) A three-dimensional representation of the measured local-trapped-field distribution. (b) The local magnetic field as a function of relative position in different radial directions, including $\theta = 0$, 0.25, 0.50, and 0.75 π, as shown in panel (c). (c) Top view of magnetic field distribution. Dashed circles represent contour lines of equivalent local magnetic fields.

**Fig. 4.** (Color online)  Time dependence of magnetic field trapped in bulk-MgB$_2$ magnet with different magnetic-flux-stability conditions. (a) The field is normalized by the remnant trapped field $B(t = 0)$ immediately after magnetization. The bulk-MgB$_2$ sample was magnetized at 20 K (orange curve) and at 30 K (pink curve) and isothermally maintained for 3 days (flux-creep state). The bulk-MgB$_2$ sample was magnetized at 20 K and maintained for 7 days at 19 K (purple curve; raw data on a magnified absolute-field scale is shown in panel (b), which is 1 K less than the magnetizing temperature (flux-frozen state). (b) Stable trapped field of ~2.91 T at 19 K for 1 week generated by circulating supercurrent in persistent current mode. Dashed lines show decay rates (1 ppm/h) of reference magnetic field.

Figure 1

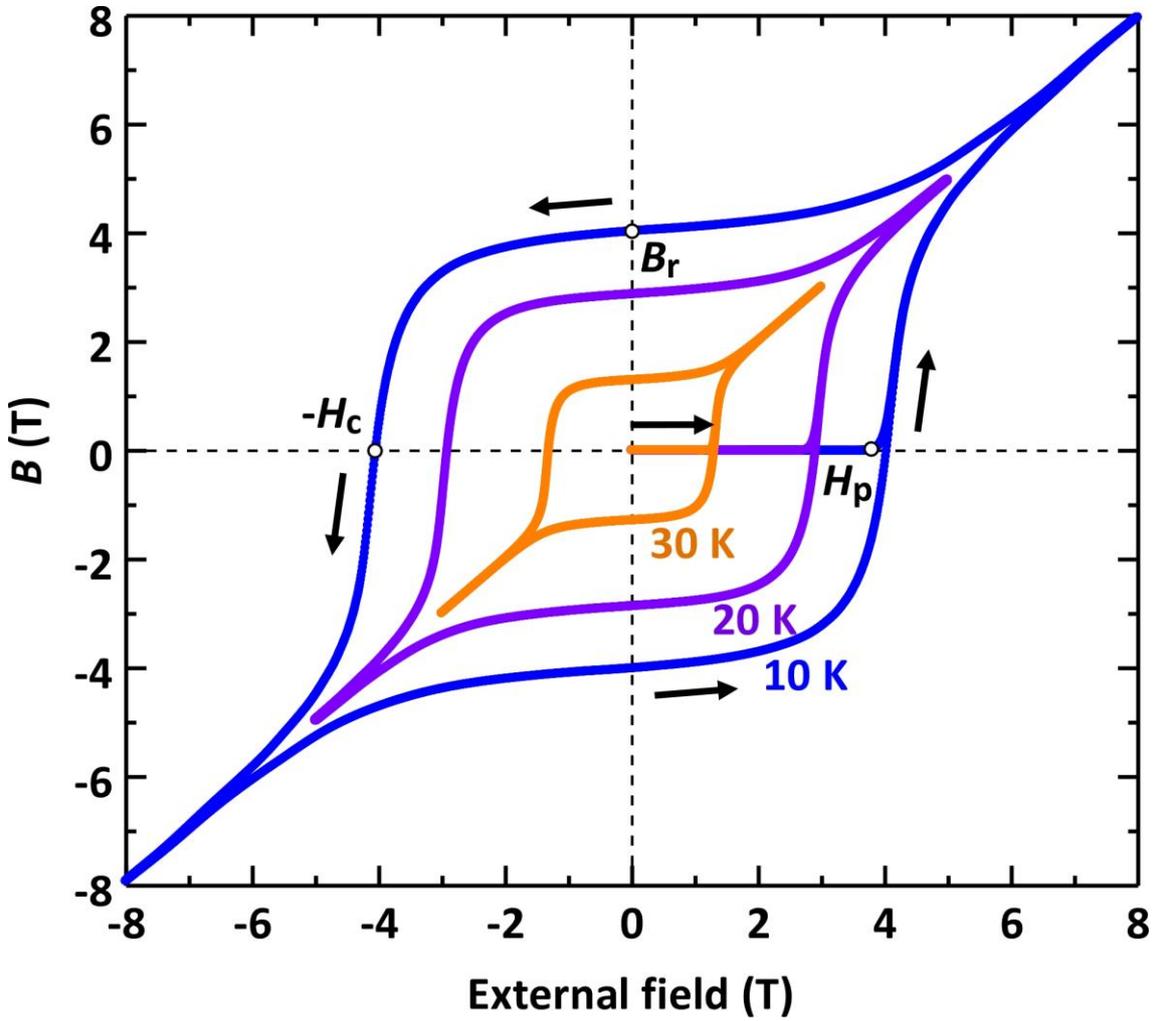

**Fig. 1.** (Color online) Magnetic hysteresis loops of bulk-MgB$_2$ magnet at 10, 20, and 30 K. The magnetic field was measured at the center of the bulk-MgB$_2$ disk pair. Arrows indicate the counterclockwise direction of the loop. $H_p$, $B_r$, and $-H_c$ represent the penetration field, the trapped remnant field, and the coercive field, respectively.

Figure 2

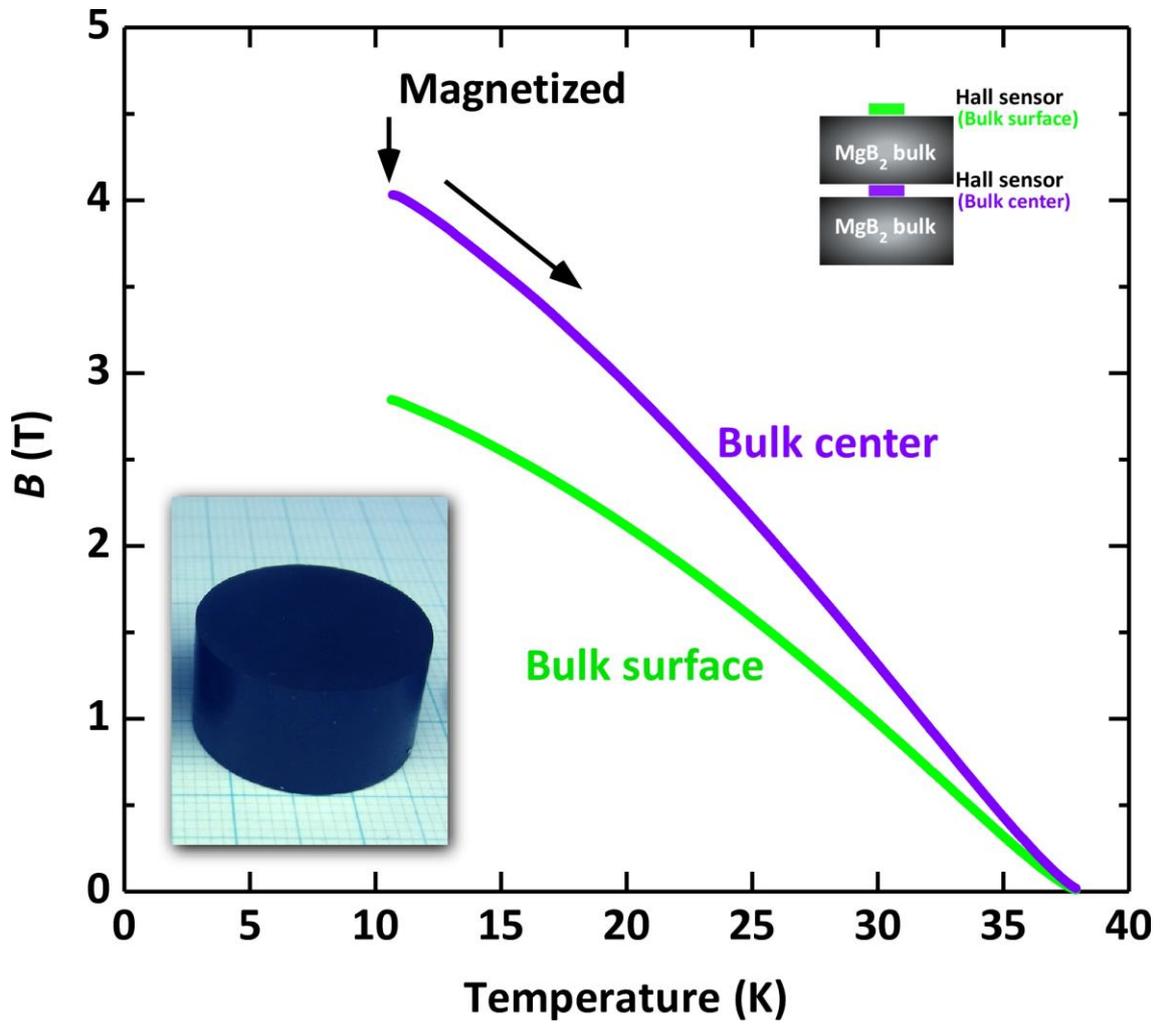

**Fig. 2.** (Color online) Trapped magnetic field measured as a function of continuously increasing temperature at center and surface of magnetized bulk-MgB$_2$ disk pair. The upper right inset shows schematic cross-sectional configuration of the bulk pair: two MgB$_2$ bulk disks and two Hall sensors. The lower-left inset shows the photograph of a bulk-MgB$_2$ disk.

Figure 3

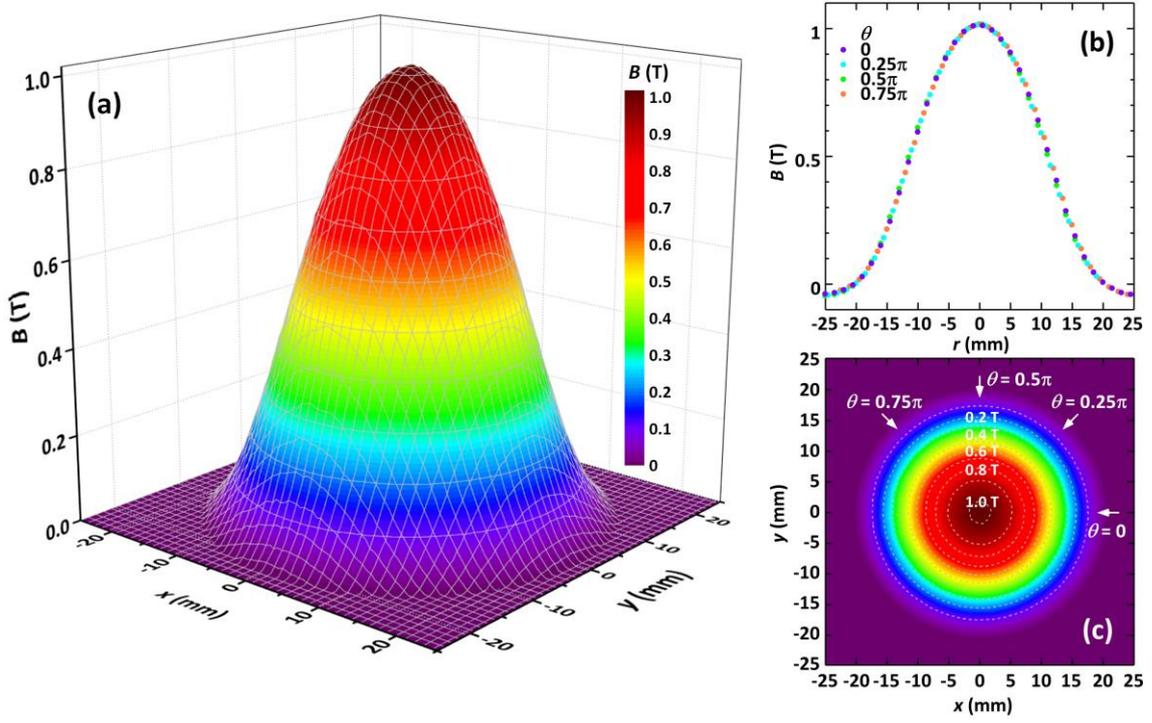

**Fig. 3.** (Color online) Distribution of trapped magnetic field in MgB$_2$ bulk disk magnet at 20 K measured 3 mm above bulk surface. (a) A three-dimensional representation of the measured local-trapped-field distribution. (b) The local magnetic field as a function of relative position in different radial directions, including $\theta$ = 0, 0.25, 0.50, and 0.75 π, as shown in panel (c). (c) Top view of magnetic field distribution. Dashed circles represent contour lines of equivalent local magnetic fields.

Figure 4

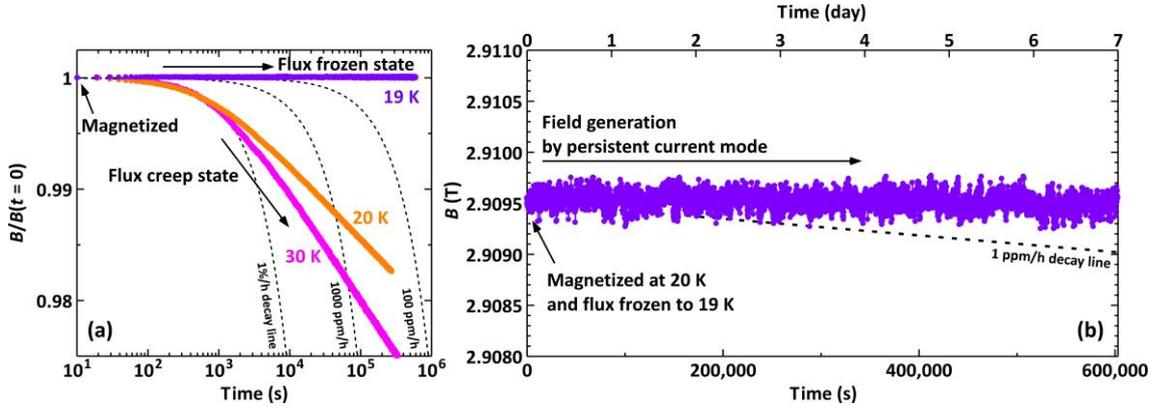

**Fig. 4.** (Color online) Time dependence of magnetic field trapped in bulk-MgB$_2$ magnet with different magnetic-flux-stability conditions. (a) The field is normalized by the remnant trapped field $B(t = 0)$ immediately after magnetization. The bulk-MgB$_2$ sample was magnetized at 20 K (orange curve) and at 30 K (pink curve) and isothermally maintained for 3 days (flux-creep state). The bulk-MgB$_2$ sample was magnetized at 20 K and maintained for 7 days at 19 K (purple curve; raw data on a magnified absolute-field scale is shown in panel (b), which is 1 K less than the magnetizing temperature (flux-frozen state). (b) Stable trapped field of ~2.91 T at 19 K for 1 week generated by circulating supercurrent in persistent current mode. Dashed lines show decay rates (1 ppm/h) of reference magnetic field.